\documentclass[11pt]{article}
\newcommand{\ben}{\begin{eqnarray}}
\newcommand{\een}{\end{eqnarray}}
\newcommand{\nnu}{\nonumber\\}

\begin{document}

\title{\bf Positivity bounds for Sivers functions}

\date{\today}

\author{Zhong-Bo Kang$^{1}$\footnote{zkang@bnl.gov} and
            Jacques Soffer $^{2}$\footnote{jacques.soffer@gmail.com}
\\[0.3cm]
{\normalsize \it $^1$RIKEN BNL Research Center, Brookhaven National Laboratory, Upton, NY 11973, USA}
\\[0.1cm]
{\normalsize \it $^2$Physics Department, Temple University, Philadelphia, PA 19122, USA}}
\maketitle

\begin{abstract}
\noindent
We generalize a positivity constraint derived initially for parity-conserving 
processes to the parity-violating ones, and 
use it to derive non-trivial bounds on several Sivers functions, 
entering in the theoretical description of single spin asymmetry for various 
processes.
\\[0.3cm]
\\[0.1cm]
\end{abstract}
\hskip 0.1cm

The relevance of positivity in spin physics, which puts non-trivial model 
independent constraints on spin observables has been already clearly demonstrated. Different methods can be used to establish these constraints and 
many interesting cases have been presented in a recent 
review article \cite{Artru:2008cp}. 
As a well known example let us remind the positivity bound on the quark 
transversity distribution $\delta{q}(x, Q^2)$ \cite{Soffer:1994ww}, 
which survives up to next-to-leading order
QCD corrections. This distribution plays a crucial role in transverse spin 
physics. In this letter we will show that one can also derive non-trivial 
positivity bounds on Sivers functions \cite{Siv90}, an important physical quantity for 
the interpretation of transverse single spin asymmetries (SSAs).
Let us first recall a powerful general result which constrains the SSA in an 
inclusive reaction of the type
$p^\uparrow+ p^\uparrow \to C + X$,
where both initial protons are transversely polarized and $C$ is any particle produced at rapidity $y$, whose polarization is not analyzed. It has been proven \cite{Soffer:2003qj} \footnote{See also Eq.~(4.25) in Ref.~\cite{Artru:2008cp}} that for $y = 0$ and
parity-conserving processes
\ben 
1 - A_{NN} (y = 0) \geq 2\left|A_N (y = 0)\right|
\label{main}
\een
where $A_{NN}$ is the double transverse spin asymmetry. In the following we will first generalize Eq.~(\ref{main}) to include parity-violating processes and show it is valid in full generality for both parity-conserving and violating ones, then use it in the case where $C$ is $W^{\pm}$ to put non-trivial bounds on the Sivers functions.

To start, we consider the inclusive reaction of the type
\ben
A(\mbox{spin 1/2})+B(\mbox{spin 1/2})\to C+X\,,
\label{ABC}
\een
where both initial spin $1/2$ particles can be in any possible directions and no polarization is observed in the final state. The observables of this reaction, which are the spin-dependent differential cross sections with respect to the momentum of $C$, can be expressed in terms of the discontinuities (with respect to the invariant mass squared of $X$) of the amplitudes for the forward three-body scattering
\ben
A+B+\overline{C}\to A+B+\overline{C}\,,
\een
as given by the generalized optical theorem. 

The spin-dependent cross section corresponding to (\ref{ABC}) can be defined through the $4\times 4$ cross section matrix $M$ and the spin density matrix $\rho$ as \cite{Artru:2008cp, Soffer:2003qj}
\ben
\sigma\left(P_a, P_b\right)={\rm Tr}\left(M\rho\right),
\label{den}
\een
where $P_a$, $P_b$ are the spin unit vectors of $A$ and $B$, $\rho=\rho_a\otimes \rho_b$ is the spin density matrix with $\rho_a=(I_2+P_a\cdot \vec{\sigma}_a)/2$, and similar for $\rho_b$. Here $I_2$ is the $2\times 2$ unit matrix, and $\sigma=(\sigma_x, \sigma_y, \sigma_z)$ stands for the $2\times 2$ Pauli matrices. The cross section matrix $M$ can be parametrized as
\ben
M&=&\sigma_0\left[I_4+A_{aN}\sigma_{ay}\otimes I_2+A_{bN}I_2\otimes \sigma_{by}
+A_{NN}\sigma_{ay}\otimes \sigma_{by}+A_{LL}\sigma_{az}\otimes \sigma_{bz}
\right.
\nnu
&&
+A_{SS}\sigma_{ax}\otimes \sigma_{bx}+A_{LS}\sigma_{az}\otimes\sigma_{bx}
+A_{SL}\sigma_{ax}\otimes\sigma_{bz}
\nnu
&&
+A_{aL}\sigma_z\otimes I_2+A_{bL}I_2\otimes \sigma_z
+A_{aS}\sigma_x\otimes I_2+A_{bS}I_2\otimes \sigma_x
\nnu
&&
\left.
+A_{LN}\sigma_z\otimes \sigma_y+A_{NL}\sigma_y\otimes \sigma_z
+A_{NS}\sigma_y\otimes \sigma_x+A_{SN}\sigma_x\otimes \sigma_y\right],
\label{matrix}
\een
where $\sigma_0$ stands for the spin-averaged cross section, and $L$, $S$, $N$ are unit vectors, in the center-of-mass system, along the incident momentum, along the normal to the scattering plane which contains $A$, $B$ and $C$, and along $N\times L$, respectively. It is important to realize that for a parity-conserving process there are only ${\it eight}$ independent spin-dependent observables 
\cite{Artru:2008cp, Soffer:2003qj, Goldstein:1975ci}: the unpolarized cross section, {\it two} single transverse spin asymmetries $A_{aN}$ and $A_{bN}$, and {\it five} double spin asymmetries $A_{NN}$, $A_{LL}$, $A_{SS}$, $A_{LS}$ and $A_{SL}$. Now for the most general case including also the parity-violating processes, one has {\it sixteen} independent spin-dependent observables. Besides those in the parity-conserving processes, one has {\it four} additional single spin asymmetries 
$A_{aL}$, $A_{bL}$, $A_{aS}$ and $A_{bS}$, and {\it four} additional double spin asymmetries $A_{LN}$, $A_{NL}$, $A_{NS}$ and $A_{SN}$.

Eq.~(\ref{matrix}) is fully justified, since substituting it back to Eq.~(\ref{den}) one has
\ben
\sigma\left(P_a, P_b\right)&=&\sigma_0[1+A_{aN}P_{ay}+A_{bN}P_{by}
+A_{NN}P_{ay}P_{by}+A_{LL}P_{az}P_{bz}
+A_{SS}P_{ax}P_{ax}
\nnu
&&
+A_{LS}P_{az}P_{bx}+A_{SL}P_{ax}P_{bz}
+A_{aL}P_{az}+A_{bL}P_{bz}
+A_{aS}P_{ax}+A_{bS}P_{bx}
\nnu
&&
+A_{LN}P_{az}P_{by}+A_{NL}P_{ay}P_{bz}
+A_{NS}P_{ay}P_{bx}+A_{SN}P_{ax}P_{by}]
\een
The crucial point is that $M$ is a Hermitian and positive matrix and in order to 
derive the positivity conditions, one should write the explicit expression of $M$ as given
by Eq.~(\ref{matrix}). In the transverse basis where $\sigma_y$ is diagonal, we have found that the diagonal matrix elements $M_{ii}$ are given by
\ben
M_{11}&=&\left(1+A_{NN}\right)+\left(A_{aN}+A_{bN}\right),
\\
M_{22}&=&\left(1-A_{NN}\right)+\left(A_{aN}-A_{bN}\right),
\\
M_{33}&=&\left(1-A_{NN}\right)-\left(A_{aN}-A_{bN}\right),
\\
M_{44}&=&\left(1+A_{NN}\right)-\left(A_{aN}+A_{bN}\right).
\een
Since one of the necessary conditions for a Hermitian matrix to be positive definite is that all the diagonal matrix elements has to be positive $M_{ii}\geq 0$, we thus derive
\ben
1\pm A_{NN}\geq \left|A_{aN}\pm A_{bN}\right|.
\label{general}
\een
Since both $A_N$ and $A_{NN}$ are parity-conserving asymmetries, it is not surprizing that this positivity bound obtained for parity-conserving processes is preserved.\\
Back to the case for $p^\uparrow+p^\uparrow\to C+X$ where the initial particles are identical, we have $A_{aN}(y)=-A_{bN}(-y)$. Using this relation in Eq.~(\ref{general}), one obtains for $y=0$,
\ben
1 - A_{NN} (y = 0) \geq 2\left|A_N (y = 0)\right|.
\label{better}
\een
We thus found Eq.~(\ref{main}) is valid in full generality, for both parity-conserving and parity-violating processes. In fact we found that physically $1-A_{NN} \pm 2 A_N$ at $y=0$ is nothing but the cross sections $\sigma (\uparrow \downarrow)/\sigma_0$ or $\sigma (\downarrow \uparrow)/\sigma_0$ at 90 degrees, which must be positive. Here $\uparrow$ and $\downarrow$ denote transversity states.

Now let us study the implication of Eq.~(\ref{main}) in a parity-violating process $p^\uparrow+p^\uparrow\to W^{\pm}+X$. The double transverse spin asymmetry $A_{NN}$ for this process has been studied in Ref.~\cite{Bourrely:1994sc, Boer:1999uu, Boer:2000er, Arnold:2008kf}. The current
conclusion is that $A_{NN}$ is expected to be negligible. The reasoning is the following:
in collinear factorization formalism, the relevant function - the transversity distribution does not contribute as pointed out in \cite{Bourrely:1994sc, Boer:1999uu}. This is due to 
the fact that the $W^{\pm}$ bosons are pure left-handed, their electroweak couplings do not allow a right-left interference. Higher-twist contributions have been studied
in Ref.~\cite{Boer:2000er}, where it was shown that a large suppression occurs from Sudakov factors, which increases with $Q$, the invariant mass. On the other hand, in the $k_T$-factorization formalism, as shown in full details in Ref.~\cite{Arnold:2008kf},
one could (potentially) have additional non-zero TMD contributions \footnote{Note in Ref.~\cite{Arnold:2008kf}, these terms are written for Drell-Yan dilepton production, not for $W$ production.}: a) contribution involving the product of $h_1$ and/or $h_{1T}^{\perp}$, and b) contribution involving $f_{1T}^{\perp}\,f_{1T}^{\perp}$ and $g_{1T}\, g_{1T}$. We want to point out the terms from (a) do not contribute for $W$ production. This is because $h_1$ and $h_{1T}^{\perp}$ have the same Dirac structure as transversity in collinear factorization, thus the same reasoning for transversity not contributing in collinear factorization applies also here. On the other hand, for the terms from (b), there could be a Sudakov suppression \cite{Boer:2001he} applies to such terms that makes the contributions very small. Even if such a Sudakov suppression does not apply, 
according to current phenomenology since $f_{1T}^\perp$ and $g_{1T}$ are much smaller than $f_1$, they are expected to generate very small effects to $A_{NN}$.
Thus we will assume $A_{NN}\approx 0$ \footnote{If $A_{NN}$ is found to be somewhat much larger than our expectation in the future, our positivity bounds could be modified accordingly using Eq.~(\ref{better}).}, and therefore Eq.~(\ref{main}) reduces to
\ben
1 \geq 2|A_N (y = 0)|\, ,
\een
to be compared with the usual trivial bound $1 \geq |A_N (y = 0)|$. There are also some studies that $A_{NN}$ could receive non-zero contribution from
the physics beyond the standard model \cite{Boer:2000er, Boer:2010mc}, but we will not discuss this in our paper. We will now see how this strong result can be used to put severe constraints on the Sivers functions.

Let us start with the exact definition of the quark Sivers function. Following the convention
of Ref.~\cite{Anselmino:2008sga}, the transverse momentum dependent (TMD) 
quark distribution in a transversely polarized hadron can be expanded as
\ben
f_{q/h^\uparrow}(x,\mathbf{k}_{\perp},\vec{S})
\equiv 
f_{q/h}(x,k_{\perp}) +  
\frac{1}{2}\Delta^N f_{q/h^\uparrow}(x,k_\perp)\,
\vec{S}\cdot \left(\hat{p}\times \hat{\mathbf{k}}_\perp \right)\,,
\een
where $\hat{p}$ and $\hat{\mathbf{k}}_\perp$ are the unit vectors of $\vec{p}$ and 
$\mathbf{k}_\perp$, respectively. $f_{q/h}(x,k_\perp)$ is the spin-averaged TMD distribution, and $\Delta^N f_{q/h^\uparrow}(x,k_\perp)$ is the Sivers function \cite{Siv90}. There is a trivial positivity bound for the Sivers functions which reads 
\footnote{For non-trivial bounds involving several different TMD distributions see \cite{Bacchetta:1999kz} and also Eq.~(4.86) in Ref.~\cite{Artru:2008cp}}
\ben
|\Delta^N f_{q/h^\uparrow}(x,k_\perp)|\leq 2f_{q/h}(x,k_{\perp}).
\label{oldbound}
\een

Let's see how the SSA of the $W$ bosons could give further non-trivial bounds 
on the Sivers functions. The SSA of the $W$ bosons in hadronic collisions 
$A^\uparrow B\to W+X$ has been studied in terms of the TMD factorization 
formalism in \cite{soffer_W, Kang:2009bp}. It is given by the ratio of the 
spin-dependent and spin-averaged cross sections,
\ben
A_N
&\equiv & \left.
\frac{d\Delta\sigma(\vec{S}_\perp)}
     {dy\,d^2\mathbf{q}_\perp} \right/
\frac{d\sigma}
     {dy\,d^2\mathbf{q}_\perp}\, ,
\een
where $y$ and $\mathbf{q}_\perp$ are the rapidity and the transverse 
momentum of the $W$ boson, respectively. The spin-dependent and 
spin-averaged cross sections are given by 
\ben
\frac{d\Delta\sigma(\vec{S}_\perp)}
     {dy\,d^2\mathbf{q}_\perp}
&=&
\frac{\sigma_0}{2}\sum_{a,b}\left|V_{ab}\right|^2
\int 
\vec{S}_\perp\cdot (\hat{p}_A\times\hat{\mathbf{k}}_{a\perp})
\Delta^N f_{a/A^\uparrow}(x_a,{k}_{a\perp}) f_{b/B}(x_b,{k}_{b\perp})\,,
\label{spin_dpt}
\\
\frac{d\sigma}{dy\,d^2\mathbf{q}_\perp}
&=&
\sigma_0 \sum_{a,b}\left|V_{ab}\right|^2
\int f_{a/A}(x_a,{k}_{a\perp})f_{b/B}(x_b,{k}_{b\perp}) \,,  
\label{spin_avg} 
\een
where the simple integral symbol represents a complicated integral, 
$\int=\int d^2\mathbf{k}_{a\perp} d^2\mathbf{k}_{b\perp}
\delta^2(\mathbf{q}_\perp-\mathbf{k}_{a\perp}-\mathbf{k}_{b\perp})$.
The partonic cross section $\sigma_0$ is given by 
$\sigma_0=(\pi/3)\sqrt{2}\,{\rm G_F}M_W^2/s$ with the 
Fermi weak coupling constant G$_F$ and $s=(p_A+p_B)^2$. 
$\sum_{ab}$ runs over all light (anti)quark flavors, $V_{ab}$ are 
the CKM matrix elements for the weak interaction. 
$x_a$ and $x_b$ are the parton momentum fractions given by
\ben
x_{a} = \frac{M_W}{\sqrt{s}}\, e^{y}, \quad
x_b = \frac{M_W}{\sqrt{s}}\, e^{-y}.
\label{xaxb}
\een

To further simplify the expression of $A_N$, it is usually assumed in the phenomenological studies that the $x$ and $k_\perp$ dependence 
of the TMD distributions can be further factorized as 
follows \cite{Anselmino:2008sga, Collins:2005rq}
\ben
f_{q/h}(x,k_\perp) 
&=& f_q(x)g(k_\perp)\,,
\\
\Delta^N f_{q/h^\uparrow}(x,k_\perp)
&=&
\Delta^N f_{q/h^\uparrow}(x)\,h(k_\perp)g(k_\perp)\, ,
\een
For the $k_\perp$ dependence, a Gaussian ansatz is 
usually introduced \cite{Anselmino:2008sga, Collins:2005rq}, 
which has the form
\ben
g(k_\perp)=\frac{1}{\pi\langle k_\perp^2\rangle}
e^{-k_\perp^2/\langle k_\perp^2\rangle}
\\
h(k_\perp)
=
\sqrt{2e}\, \frac{k_\perp}{M_1}
e^{-k_\perp^2/M_1^2}
\een
The smart choice of $h(k_\perp)$ is introduced in 
\cite{Anselmino:2008sga}, which satisfies $h(k_\perp)\leq 1$. 
Thus the positivity bound in Eq.~(\ref{oldbound}) implies
\ben
\left|\Delta^N f_{q/h^\uparrow}(x)\right|\leq  2f_q(x).
\een
With the Gaussian ansatz, one could carry out the integration 
in Eqs.~(\ref{spin_dpt}) and (\ref{spin_avg}) analytically and obtains
\ben
A_N(y=0) &=& H(q_\perp)
\frac{\sum_{ab}|V_{ab}|^2 
      \Delta^N f_{a/A^\uparrow}(x)\,f_{b}(x)}
     {\sum_{ab}|V_{ab}|^2\,f_{a}(x)\, f_{b}(x)} ,
\label{An_W}
\een
where $x=M_W/\sqrt{s}$ for $y=0$ from Eq.~(\ref{xaxb}), and 
$H(q_\perp)$ is given by
\ben
H(q_\perp)=\vec{S}_\perp \cdot (\hat{p}_A\times \mathbf{q}_\perp)\,
\frac{\sqrt{2e}}{M_1}
\frac{\langle k_s^2\rangle^2}
     {[\langle k_\perp^2\rangle + \langle k_s^2\rangle]^2}\,
     e^{-\left[ \frac{\langle k_\perp^2\rangle-\langle k_s^2\rangle}
     {\langle k_\perp^2\rangle + \langle k_s^2\rangle}\right] 
     \frac{\mathbf{q}_\perp^2}{2\langle k_\perp^2\rangle}},
\een
where $\langle k_s^2\rangle=M_1^2\, \langle k_\perp^2\rangle
/[M_1^2+\langle k_\perp^2\rangle]$.
The $q_\perp$-dependent function $H(q_\perp)$ reaches its 
maximum $H(q_\perp)_{\rm max}$ when 
$q_\perp^2=\langle k_\perp^2\rangle(\langle k_\perp^2\rangle + \langle k_s^2)
/(\langle k_\perp^2\rangle - \langle k_s^2\rangle)$, 
with $H(q_\perp)_{\rm max}$ given by
\ben
H(q_\perp)_{\rm max}=\frac{\langle k_s^2\rangle^2}
     {[\langle k_\perp^2\rangle + \langle k_s^2\rangle]^2}
     \left[\frac{2\langle k_\perp^2\rangle}{M_1^2}
     \frac{\langle k_\perp^2\rangle + \langle k_s^2\rangle}
     {\langle k_\perp^2\rangle - \langle k_s^2\rangle}\right]^{\frac{1}{2}}.
\een
Using the fact that $1\geq 2|A_N (y = 0)|$ for any $q_\perp$ and 
$\sqrt{s}$ in Eq.~(\ref{An_W}), we thus derive a new bound 
for the Sivers functions
\ben
\frac{\left|\sum_{ab}|V_{ab}|^2 
      \Delta^N f_{a/A^\uparrow}(x)\,f_{b}(x)\right|}
     {\sum_{ab}|V_{ab}|^2\,f_{a}(x)\, f_{b}(x)}\leq 
     \frac{1/2}{H(q_\perp)_{\rm max}}.
\label{newbound}     
\een
For $W^+$, Eq.~(\ref{newbound}) can be simplified as
\ben
\left|\frac{\Delta^N u(x)}{u(x)}+\frac{\Delta^N \bar{d}(x)}{\bar{d}(x)}\right|
\leq \frac{1}{H(q_\perp)_{\rm max}}.
\label{wp}    
\een
While for $W^-$, one obtains the following constraint
\ben
\left|\frac{\Delta^N d(x)}{d(x)}+\frac{\Delta^N \bar{u}(x)}{\bar{u}(x)}\right|
\leq \frac{1}{H(q_\perp)_{\rm max}}.
\label{wm}
\een
Eqs.~(\ref{newbound})-(\ref{wm}) are the main results of our paper \footnote{There are some sign errors in Ref.~\cite{soffer_W}, which have been
corrected in the l.h.s of Eqs.~(\ref{wp}, \ref{wm}).}.
Even though we derive the bounds for the Sivers functions in $W^{\pm}$ production in $pp$ collisions (or Drell-Yan (DY) type process), our results also apply to the Sivers functions in semi-inclusive deep inelastic scattering (SIDIS) process. This is because one could show from the parity and time-reversal invariance that the Sivers function in SIDIS and in DY have the same functional form but opposite signs \cite{Collins:2002kn}.

If we further use the values for the parameters $M_1^2$ and 
$\langle k_\perp^2\rangle$ from the global fitting of the 
experimental data for the SSA of the hadron in 
SIDIS processes, 
say, from Ref.~\cite{Anselmino:2008sga}, 
\ben
\langle k_\perp^2\rangle=0.25~{\rm GeV}^2 \qquad M_1^2=0.34~{\rm GeV}^2
\een
one obtains $H(q_\perp)_{\rm max}\approx 0.31$, 
thus $1/H(q_\perp)_{\rm max}\approx 3.2$. The fitted value of $M_1^2$ in Ref.~\cite{Anselmino:2008sga} is $M_1^2=0.34^{+0.30}_{-0.16}$. If one uses $M_1^2 = 0.64$, we get a better bound, 2.6 instead of 3.2. And at the same time the Sivers functions get bigger as shown in Ref.~\cite{Anselmino:2008sga}. The constraints become better. While the previous 
bound in Eq.~(\ref{oldbound}) would give the number 4 in 
the right hand side of the inequalities in Eqs.~(\ref{wp}) 
and (\ref{wm}), thus the new bound is slightly stronger 
than the previous one. However if one uses the lower value $M_1^2 = 0.18$, the bound becomes 4.75, which 
is larger than the trivial one. Also if one increases $\langle k_T^2\rangle$ while keeping the same $M_1^2$, the bound will become looser. For example, for 
$M_1^2=0.64$, when $\langle k_T^2\rangle>0.75~{\rm GeV}^2$, then the bound will become looser than the trivial one. 

To summarize, in the framework of the factorized Gaussian ansatz, we have derived non-trivial bounds for the Sivers functions 
associated with $u$, $d$, $\bar{u}$, and $\bar{d}$ quarks. 
The bound is slightly stronger than the previously derived 
one on the Sivers functions. Our results impose important 
constraints in the global analysis of the Sivers functions 
in the phenomenological studies of the 
single transverse spin asymmetry. 
We emphasize that the significance of our results should 
not be restricted to one special parametrization
based on a very limited set of data. It is also important to 
realize that our newly-derived 
bounds Eqs.~(\ref{newbound})-(\ref{wm}) can be used as a consistent 
check of the quark Sivers functions extracted from the global analysis
of the single transverse spin asymmetries.

Finally this also applies to the gluon Sivers function which is accessible in the SSA for direct photon production \cite{Schmidt:2005gv}, because
in this case also it was found that $A_{NN}$ is very small \cite{Soffer:2002tf, Mukherjee:2003pf}, say hardly more than 1 or 2\%. The most appropriate reaction is in fact $pp^{\uparrow}\rightarrow \gamma + \mbox{jet} + X$, which will deserve further considerations. We look forward to future accurate measurements of single spin asymmetries to greatly
improve our knowledge about Sivers functions.
\\[0.5cm]
J.S. acknowledges some interesting discussions with X. Artru. 
We thank A. Metz for helpful discussions and useful comments. 
Z.K. is grateful to RIKEN, Brookhaven National Laboratory, 
and the U.S. Department of Energy (Contract No.~DE-AC02-98CH10886) 
for supporting this work.

%%%%%%%%%%%%%%

\end{document}